\documentstyle[12pt]{article}

\newcommand{\tr}{\mathop{\rm tr}\nolimits}
\normalbaselines
\catcode `\@=11
\@addtoreset{equation}{section}

\def\section{\@startsection {section}{1}{\z@}{-3.5ex plus -1ex minus
     -.2ex}{2.3ex plus .2ex}{\normalsize\bf}}
\def\subsection{\@startsection{subsection}{2}{\z@}{-3.25ex plus -1ex minus
-.2ex}{1.5ex plus .2ex}{\normalsize\bf}}
\def\thebibliography#1{\section*{References}
\list
  {[\arabic{enumi}]}{\settowidth\labelwidth{[#1]}\leftmargin\labelwidth
  \advance\leftmargin\labelsep
  \usecounter{enumi}}
  \def\newblock{\hskip .11em plus .33em minus -.07em}
  \sloppy
  \sfcode`\.=1000\relax}

\catcode `\@=12

\begin{document}

\vspace*{2.5cm}

\begin{center}

{\bf NEUTRINOS OF UNIVERSE AND MASSES OF PARTICLES}\vspace{1.3cm}\\
\medskip

{\bf Valery Koryukin}
\vspace{0.3cm}\\

 Department of Applied Mathematics, Mari State Technical University\\
 a/b 179, Main Post--Office, Yoshkar--Ola, 424000, Russia

\end{center}

\vspace*{0.5cm}

{\small
\noindent

 The detection and the research of the neutrinos background of
 Universe are the attractive problems. This problems do not
 seem the unpromising one in the case of the high neutrinos
 density that is necessary for the explanation of the
 nucleons-antinucleons asymmetry of Universe (the Pontekorvo's
 and Smorodinsky's hipothesis~\cite{PS}).

 It was offered before to use the low energy neutrinos background
 of Universe for the explanation of the gravitational phenomena
 with the quantum position attracting the Casimir's  effect for
 this~\cite{K}. As a result it was connected the gravitational
 constant~$G_N$ with the parameters characterizing the electroweak
 interactions: $G_N=\sigma /6$, where~$\sigma$ is the scattering
 cross-section of neutrinos upon the particle of the macroscopic body.
 In particular (by the employment of the neutrinos distribution with
 the effective temperature~$T=1.9 K$) the average value~$<\sigma >$
 of the scattering cross-section will writte down as
\begin{equation}
 <\sigma > \approx \frac{4}{\pi} k
 G_F^2 <\omega>^2 (\frac14 - \frac12 \xi + \xi^2) ,
\end{equation}
 where~$G_F$ is the Fermi's constant,
 $\xi = \sin^2\Theta_W$ ($\Theta_W$ is the weak angle), the
 average energy of neutrinos $<\omega> \approx 3.15 T$.
 The factor~$k$ connects with the collision radiation and for the
 electron it is directly proportional ($k \approx (2/3)\alpha$) to
 the fine structure constant~$\alpha \approx 1/137$.

 If now we shall be based on the results of the experements fixing
 the equality of the gravitation mass and the inert one then it can
 consider that the spectrum of the particle masses is defined by
 their interaction with the neutrinos background of Universe. This
 statement is confirmed what the rest mass of the photon is equal to
 zero in contradistinction to the masses of the vector
 bosons $W^{+}$, $W^{-}$, $Z^{0}$ whiches interact with the neutrinos
 immediately~\cite{K1}.}

\section{\hspace{-4mm}.\hspace{2mm} THE SPACE OF THE
 WAVE--FUNCTIONS WITH THE SEMISCALAR PRODUCT}

 As is known the many theoretical models of the interaction particles
 do not without the scalar fields. At the same time the fundamental
 (uncompound) scalar particles are not detected by experiment and the
 theorists invent the different method to explain this unpleasant
 fact. We choose the other way, constracting the theory in which the
 fundamental scalar fields are absent and all problems are solved at
 the expense of the particles having the spin.

 One of the fundamental problems of the
 theoretical physics is the search of the axioms, which ought
 to be the basis for the one--valued constraction of
 Lagrangians of the relativistic fields. The creation of the
 gauge fields theory was the great success in the solution of
 this problem. The gauge formalism allowed to get the total
 Lagrangians from the Lagrangians for the noninteracting (free)
 fields and in the first place by the extension of their partial
 derivatives. In consequence of this with all urgency the task
 arised to find the principles of the constraction of
 Lagrangians of the free fields. But it is unlikely the free
 fields can exist, and if even they exist, then it is unlikely
 we can discover them. Therefore, it is logically to get the
 Lagrangians of the free fields from the more general ones in
 the certain approximation, in consequence of this we must
 seek the principles of the constraction of the total
 Lagrangians for the interaction fields all the same.

 If the spinor fields be absented, one of this principles was
 able be the demand, that the Lagrangian is the quadratic one
 in the derivatives (if only in the first approximation).
 Note, that Skyrme~\cite{sky} considered the fermions as the
 convenient means of the mathematical descreption only. It was
 insisting him to conduct the search of the method of the
 construction of the fermion states from the boson fields
 allowed to turn de Broglie-Heisenberg's scheme of the
 confluence (the boson from the fermions). Exactly in
 consequence of this, we make the attempt to get the Dirac's
 equatios from the quadratic Lagrangian in the derivatives of
 the fields.

 Considering an arbitrary quantum ensemble we shall introduce the set
 of $N$--component ``empiric'' functions~$\Psi (\omega)$ for its
 description. It is convenient to consider the functions~$\Psi (\omega)$
 depending on the~$r$ parameters~$\omega^a $, as the coordinates of the
 point~$\omega$ belonging to the ``empiric'' manifold~$M_r$. Also we
 shall introduce differentiable manifold~$M_n$ (it is possible the
 manifold~$M_n$ is connected with the macroscopic observer), which it
 make sence to call the ``theoretical'' one and it is possible which is
 a submanifold in the manifold~$M_r$. As a result the every
 field~$\Psi (\omega)$ induces the set of the~$N$--component
 ``theoretical'' functions~$\Psi (x)$ ($x\in M_n$), the equations of
 which are being obtained from the demand of the minimality of the
 generalized variance of the fields~$\Psi (x)$.

 So, we shall consider the quantum
 system for a description of which we shall use the wave--functions
 belonging to the vector functional space with semiscalar product.
 As in the general case the functions~$\Phi, \Psi, \Theta,...$
 will belong to the complex vector space~$L$, then the complex--valued
 function~$<\!\Phi, \Psi \! >$, being semiscalar product, must
 satisfy to the following conditions
$$
     1^o) <\!\Phi, \Psi \! > = <\! \Psi, \Phi \! >^* ;
$$
$$
     2^o) <\! \lambda \Phi+\nu \Psi,\Theta \! > =
          \lambda <\! \Phi, \Theta\! > + \nu <\!\Psi, \Theta\! > ;
$$
$$
     3^o) <\!\Psi, \Psi\! > \; \geš0 .
$$
 ÇÄÅ~$*$ is the symbol of the complex conjugation, $\lambda$ and~$\nu$
 are the complex numbers. Certainly, the Hilbertian space for the
 description of the quantum systems is used, but we wish to rule out
 the axiom in consequence of which only the nullvector satisfies to
 the condition
$$
     4^o) <\!\Psi, \Psi\! > = 0 .
$$
 Of course, not all the wave--functions can have the probability sense
 in this case, but we agree to this consciously so as to have the
 possibility to describe the generalized coherent states~\cite{P}.
 By this one of the main conditions of the set of the wave--functions
 will be absent this is the possibility of the orthogonalization of
 them.

 We shall rely on the approach
 suggested by Schr\"odinger~\cite{S} which introduced
 the set of the unorthogonal to each other wave functions
 describing the unspreading wave packet for a quantum
 oscillator. Later Glauber~\cite{G} showed a scope for a
 description of coherent phenomena in the optics by the
 Schr\"odinger introduced states and it was he who
 called them as coherent. This approach received the further
 development in Perelomov's work's who proposed the
 definition of the generalized coherent states specifically
 as the states arising by the action of the representation
 operator of a some transformation group on any fixed vector
 in the space of this representation~\cite{P}.

 It is what allow to give the physical interpretation to the gauge
 transformations by our opinion as the transformations inducing
 the generalized coherent states, which are characterized by the
 continuous parameters~\cite{Sh}. Admitedly if the parameters space
 are not the compact one (we shall consider the space--time manifold
 always as its subspace) then by the rather large changes of parameters
 it is necessary take into account the speed finity of the information
 propagation in consequence of what the coherence of the
 states are able to lose (what lead to the absence of the quantum
 phenomena on the macroscopic level). It makes us change the
 Perelomov's definition considering it taken place for the arbitrary
 group only in the neighbourhood of identity, what gives rise to
 generalize the given definition not only for the Lie local groups but
 and for the Lie local loops.

 As known in the theory of the scattering the initial states of
 particles are being described by the vector~$\mid\! \Psi_{in}\! >$,
 relating to the infinitely distant past and the final states are
 being described by the vector~$\mid\! \Psi_{out}\! >$, relating to
 the infinitely distant future, to in both case we have the
 justification to ignore an interaction between particles. As a
 result~$S$--matrix of the scattering is being defined by the
 relation $\mid\! \Psi_{out}\! >=S \mid\! \Psi_{in}\! >$,
 that is to say the process of the collision is being considered as
 ``black box'', described by~$S$--matrix, which transform~$in$--state
 to ~$out$--state of the system. As there are interesting the
 transitions only between the different states then it is necessary
 to subtract the unit operator~$I$ from~$S$--matrix, thereby defining
 the operator of the transition as~$T = S - I$ or in a different way
\begin{equation}
 \label{1}
     T\mid\! \Psi_{in}\! > = \mid\! \Psi_{out}\! >
 - \mid\! \Psi_{in}\! > .
\end{equation}

 Naturally, by this the space and the time are being thought of as
 trivial objects properties of which are known. But we shall
 consider that the properties are being established by devices
 having the limited possibilities and they are being analysed by an
 observer possibilities of which are not unbounded ones also.
 Moreover we shall suppose that the considered interactions are the
 littles in consequence of what it can consider that
 operator~$\{ T\}$ are the infinitesimal ones.

\section{\hspace{-4mm}.\hspace{2mm} THE LIE LOCAL LOOP}

 Let us to consider the wave packet the equivalence relation for the
 functions~$\Psi (\omega )$ of which we shall give by the
 infinitesimal transformations
\begin{equation}
 \label{1}
 \Psi \longmapsto \Psi + \delta \Psi =
 \Psi + \delta T(\Psi ) ,
\end{equation}
 where~$\delta T$ is the particular case of the transition operator
 (at the begining the symmetry type  is not being specified)
 and~$\omega $ is a set of parameters characterizing the generalized
 coherent state. Further we shall not separate the functions
 describing $in$--state and $out$--state, which are able to turn out
 to be the coherent ones in the cause--related region.

 Let~$M$ is the topological space of the parameters~$\omega $, the
 properties of which (including the dimensionality) for the present
 are not being limited and let some set of the smooth curves passes over
 the point~$\omega_o \in M$. The given set allows to produce the set of
 the tangent vector fields among which we shall the infinitesimal
 fields~$\{ \delta \xi (\omega )\}$, so as to define the deviations
 of the fields~$\Psi (\omega )$ in point~$\omega_o \in M$ as
\begin{equation}
 \label{2}
 \delta_o \Psi = \delta X(\Psi ) =
 \delta T(\Psi ) - \delta \xi (\Psi ) .
\end{equation}
 The unspreading wave packets interest us, therefore it is desirable to
 demand so as the deviations were minimal ones even if ``on the
 average'' for what it is necessary to define the
 density~$\rho = \rho (\omega )$ of the set of the
 fields~$\{ \delta X(\Psi )\}$~\cite{K2}. We shall set that the
 density~$\rho (\omega )$ induces the dimensionality of the manifold
 which contains the points of the smooth curves, even if in some
 neighbourhood of a point~$\omega_o$, and which we
 shall designate as~$M_n$. Of course in this case the dimensionality
 of the linear shell, pulled on functions~$\delta T(\Psi )$, can not be
 smaller of~$n$.

 As the experiments on the scattering of particles in which the laws
 of the conservation are being prescribed are the sole source of the
 information about the structure of the space--time manifold on the
 microscopic level taking into account the Noether's theorem we
 introduce the finite--dimensional manifold~$M_r$ of parameters~$\omega^a$
 ($a,b,c,d,e=1,2,...,r$) even if as the subspace in the considered
 space~$M$ before connecting its dimensionality~$r$ with the numbers
 of the conserved dynamical invariants. Further we shall consider
 the space~$M_r$ as the manifold, the parameters~$\omega^a$ as the
 coordinates of the point~$\omega \in M_r$ and we shall give the
 fields~$\Psi (\omega )$ in a certain domain~$\Omega_r$ of the given
 manifold.
 We choose the arbitrary point~$e$ in the domain~$\Omega_r$
 and we shall consider that this point~$e$ is the centre of the
 coordinate system. Furter we shall consider that the
 domain~$\Omega_r$ contain the subdomain~$\Omega_n$ with the
 point~$e$ by this the domain~$\Omega_n$ belong to a certain
 differentiable manifold~$M_n$ (although it is possible in is
 convenient to define the manifold~$M_n$ separately from the
 manifold~$M_r$). We shall run out the set of the smooth
 curves belonging to the manifold~$M_n$ and having the
 general point~$e$. Furter we shall define the set of the vector
 fields~${\xi (x)}$ being tangents to this curves and we shall
 consider that a point~$x \in \Omega_n$ and on the domain~$\Omega_n$
 the own coordinate system is defined. It is convenient to give the
 differentiable manifold~$M_n$ by the differentiable
 functions~$\omega^a = \omega^a (x)$ in this case.

 Let~$\delta \Omega_r$ is the sufficiently small neighbourhood of the
 point~$e$, thereby and the sufficiently small
 neighbourhood~$\delta \Omega_n$ of the point~$x$ is being given
 ($x\equiv e\in \delta \Omega_n \subset \delta \Omega_r$). The
 coordinates of the point~$x$ note as~$x^i$ ($i,j,k,l,p,q=1,2,...,n$).
 Further we shall consider the fields~$\Psi (x)$ as the cross section
 of the vector fiber bundle~$E_{n+N}$. Using the vector
 fields~$\delta \xi (x)$ the coordinates of the neighbouring
 point~$x'=x+\delta x \in \delta \Omega_n$ are being written down as
\begin{equation}
 \label{3}
 x'^i=x^i+\delta x^i \cong x^i+\delta \omega^a \xi_a^i (x) .
\end{equation}
 Comparing the values of the fields~$\Psi '(x)$ and~$\Psi (x'$), where
\begin{equation}
 \label{4}
 \Psi '(x) = \Psi + \delta\Psi = \Psi + \delta T(\Psi ) \cong
 \Psi + \delta\omega^a T_a(\Psi ) ,
\end{equation}
\begin{equation}
 \label{5}
 \Psi (x') =\Psi (x+\delta x) \cong \Psi +
 \delta \omega^a \xi_a^i \partial_i \Psi
\end{equation}
 ($\partial_i $ are the partial derivatives), we see that they are
 differing by the observables
\begin{equation}
 \label{6}
 \delta_o\Psi (x) \cong \delta\omega^a X_a(\Psi ) =
 \delta\omega^a [T_a(\Psi ) - \xi_a^i \partial_i\Psi ] ,
\end{equation}
 which can interpret as the deviations the field~$\Psi (x)$, received
 with the help of the transformations~(2.4). Further we shall consider
 the domain~$\delta \Omega_r \subset M_r$ as the domain of the Lie
 local loop~$G_r$ (specifically which can have and the structure of
 the Lie local group if we provide it with the property of the
 associativity) with the unit~$e$ induced by the set~$\{ T\}$,
 by this we shall consider the expression of the form~(2.4) as the
 infinitesimal law of the transformations of the Lie local loop of the
 fields~$\Psi (x)$. Precisely the structure of the Lie local loop will
 characterize the degree of the coherence considered by us the quantum
 system. By this the maximal degree is being reached for the Lie
 simple group and the minimal degree is being reached for the Abelian
 one. In the last case we shall have the not coherent mixture of the
 wave--functions, it's unlikely which can describe the unspreading
 wave packet that is being confirmed by the absence of the fundamental
 scalar particles, if hipothetical particles are not being taken into
 account (in experiments only the mesons, composed from the quarks,
 are being observed and which are not being considered the fundamental
 one).

 As it's unlikely it can be to ignore an interaction between particles
 we must be able to select those interactions which interest us.
 Precisely therefore it makes sense to select the set of the operators
 which will play the role of the connection in further. Demand
 that the transformations of the Lie local loop are the covariant ones
 in a point it is necessary to consider the fields
\begin{equation}
 \label{7}
 L_a(\Psi ) = T_a(\Psi )+\xi_a^i \Gamma_i \Psi
\end{equation}
 as the cross sections of those fiber bundle~$E_{n+N}$ that and
 fields~$\Psi (x)$. In consequence of this the formula~(2.6) is being
 rewrited so
\begin{equation}
 \label{8}
 \delta_o\Psi \cong \delta\omega^a X_a(\Psi ) =
 \delta\omega^a [L_a(\Psi ) - \xi_a^i \nabla_i\Psi ] ,
\end{equation}
 where~$\nabla_i$ are the covariant derivatives with respect the
 connection~$\Gamma_i (x)$. Note, if~$L_a(\Psi ) = L_a\Psi $, then the
 following relations~\cite{K3}
\begin{equation}
 \label{9}
 \xi_a^i~\nabla_i\xi_b^k - \xi_b^i~\nabla_i\xi_a^k
 - 2~S_{ij}^k~\xi_a^i~\xi_b^j = - C_{ab}^c~\xi_c^k,
\end{equation}
\begin{equation}
 \label{10}
 L_aL_b - L_bL_a - \xi_a^i~\nabla_iL_a + \xi_b^i~\nabla_iL_a
 + R_{ij}~\xi_a^i~\xi_b^j = C_{ab}^c~L_c
\end{equation}
 must take place, where ~$S_{ij}^k(x)$ are the components of the
 torsion of the space--time~$M_n$
\begin{equation}
 \label{11}
 S_{ij}^k = (\Gamma_{ij}^k - \Gamma_{ji}^k)/2
\end{equation}
 and~$R_{ij}(x)$ are the components
 of the curvature of the connection~$\Gamma_i(x)$
\begin{equation}
 \label{12}
 R_{ij} = \partial_i \Gamma_j - \partial_j \Gamma_i +
 \Gamma_i \Gamma_j - \Gamma_j \Gamma_i .
\end{equation}
 By this the components~$C_{ab}^c(x)$ of the structural tensor
 of the Lie local loop~$G_r$ must satisfy to the identities
\begin{equation}
 \label{13}
 C_{ab}^c + C_{ba}^c = 0,
\end{equation}
\begin{equation}
 \label{14}
 C_{[ab}^d~C_{c]d}^e - \xi_{[a}^i~\nabla_{|i|} C_{bc]}^e
 + R_{ij[a}^e~\xi_b^i~\xi_{c]}^j = 0,
\end{equation}
 where~$R_{ija}^e(x)$ are the components of the curvature of the
 connection~$\Gamma_{ia}^b(x)$
\begin{equation}
 \label{15}
 R_{ijb}^a = \partial_i \Gamma_{jb}^a - \partial_j \Gamma_{ib}^a +
 \Gamma_{ic}^a \Gamma_{jb}^c - \Gamma_{jc}^a \Gamma_{ib}^c .
\end{equation}

 We construct the differentiable manifold~$M_n$, not interpreting it
 by physically. Of course we would like to consider the manifold~$M_n$
 as the space--time~$M_4$. At the same time it is impossible to take
 into account the possibility of the phase transition of a system as a
 result of which it can expect the appearance of the coherent states.
 In consequence of this it is convenient do not fix the dimensionality
 of the manifold~$M_n$. It can consider that the macroscopic system
 reach the precisely such state by the collapse. As a result we have
 the classical analog of the coherent state of the quantum system.
 Besides there is the enough developed apparatus --- the dimensional
 regularization using the spaces with the changing dimensionality and
 representing if only on the microscopic level.

\section{\hspace{-4mm}.\hspace{2mm} THE GAUGE FIELDS}

 We shall demand the minimality of the variations~(2.6) (or~(2.8)), if
 only on the ``average'', in order to can be hope for the set of the
 fields~$\Psi (x)$ is capable to describe the unspreading wave packet.
 Consider for this the following integral
\begin{equation}
\label{1}
 {\cal A} = \int\limits_{\Omega_n} {\cal L} d_nV =
 \int\limits_{\Omega_n} \kappa \overline{X^b} (\Psi)
 \rho_b^a X_a (\Psi) d_n V ,
\end{equation}
 being the analogue of the fields~$\Psi (x)$ variance in the
 domain~$\Omega_n$ at issue, which we shall call the action,
 and~${\cal L}$ we shall call the Lagrangian. Here and further
 $\rho_a^b (x)$ are the components of the density matrix~$\rho (x)$
 (note that Latin indexis are the only (possible) visible part of the
 indexis of the density matrix,  $\tr \rho = 1, \rho^+ = \rho$, the
 top index~$+$ is the symbol of the Hermitian conjugation), and the
 bar means the Dirac conjugation which is the
 superposition of the Hermitian conjugation and the space inversion.
 Solutions~$\Psi (x)$ (and even one solution) of equations, which are
 being produced by the requirement of the minimality of the
 integral~(3.1) can be used for the constraction of the all set of the
 functions~$\{ \Psi (x)\}$ (generated by the transition operator),
 describing the unspreading wave packet. It is naturally
 to demand the invariance of the integral~(3.1) relatively the
 transformations~(2.3) and~(2.4), in consequence of what it is
 necessary to introduce the additional fields~$B(x)$ with the
 transformation law in point~$x \in \delta \Omega_n $ in the form
\begin{equation}
\label{2}
 \delta_o B = \delta\omega^a Y_a(B) +
 \nabla_i\delta\omega^a Z_a^i(B) ,
\end{equation}
 and which we shall name gauge ones. Make it in the standard manner
 defining them by the density matrix~$\rho (x)$ as
\begin{equation}
\label{3}
 B^b \overline{B_a} = \rho_a^b (B^c \overline{B_c}) .
\end{equation}
 Note that the factorization of the gauge fields~$B^a(x)$ on
 equivalence classes is allowed for the writing of their indexes.
 This constraction method of the gauge fields theory allows to
 remove the homogeneous background (of particles and fields)
 from the consideration and allows to consider those fluctuations
 of which structure give them to survive a enough long time on the
 temporal scale of the observer.

 Further we shall assume that the density matrix~$\rho (x)$ defines
 the dimensionality of manifold~$M_n$, using even if for this the
 corresponding generalized (singular) functions in consequence of
 what the rank
 of the density matrix~$\rho (x)$ must be equal to~$n$, and the
 formula~(3.1) can be rewrited in the form
\begin{equation}
\label{4}
 {\cal A} = \int\limits_{\Omega_r} {\cal L} d_rV =
 \int\limits_{\Omega_r} \kappa \overline{X^b} (\Psi)
 \rho_b^a X_a (\Psi ) d_r V ,
\end{equation}
 We would connect the rank of the density matrix~$\rho (x)$ with
 the nonzero vacuum average of the gauge fields.

 Introduce the metric in the space~$M_n$, using the reduced density
 matrix~$\rho '(x)$, the components of which are being note as
\begin{equation}
\label{5}
 \rho_i^j = \overline{\xi_i^b}\rho_b^a\xi_a^j /
 (\overline{\xi_k^d}\rho_d^c\xi_c^k) .
\end{equation}
 Let the fields
\begin{equation}
\label{6}
 g^{ij} = \eta^{ik} (\rho_k^j) (\eta_{pq} g^{pq})
\end{equation}
 (where~$\eta^{ij}$ are the contravariant components of the metric
 tensor of the tangent space to~$M_n$) are the components of the
 inverse tensor to the fundamental one of the Weyl space~$M_n$. As a
 result the construction of the differentiable manifold~$M_n$ can be
 connected with the finding of the equations solutions of the gauge
 fields~$\Phi^i = B^a \xi_a^i $ received from the demand of the
 minimality of the total generalized variance
\begin{equation}
\label{7}
 {\cal A}_t = \int\limits_{\Omega_n} {\cal L}_t d_nV =
 \int\limits_{\Omega_n} [\kappa \overline{X^b} (\Psi)
 \rho_b^a X_a (\Psi) + \kappa_1 (\overline{Y^b} (B)
 \rho_1{}_b{}^a Y_a (B))] d_n V .
\end{equation}

 It is naturally that the supposition about the fields filling the
 Universe and defining the geometrical structure of
 the space--time manifold, allow to introduce the connection of the
 fundamental tensor of this manifold with that kind of the statistical
 characteristic as the entropy defining it in a standard manner by the
 reduced density matrix~$\rho' (x)$ in the form
\begin{equation}
\label{8}
  S = - \tr (\rho' \ln \rho' ) .
\end{equation}
 As a result the transition from the singular state of the Universe to
 the modern one can be connected with the increase of the entropy~$S$
 defined here.

\section{\hspace{-4mm}.\hspace{2mm} THE CPT--PARITY AND THE NEUTRINOS}

 As is known~\cite{lud},~\cite{pau},~\cite{sw}
 Lorentz-invariance of the local field theory induces also
 and its óòô--invariance, in consequence of this we connect the
 existence of the fermions with the CPT--theorem having the place for
 the physical models. We assume that the spin of the particles
 characterizes those degree of freedom which induces the transitions
 between two coherent states connected by the CPT--parity operator.

 Let's assume that Universe had a stage of a development
 during which CPT-invariance of physical processes are absent.
 Such situation must arise by a break down of Lorentz invariance
 of the space-time $M_4$ in which only the fields with the
 elemental structure (scalar fields) can take place.
 This stage of a development of Universe, if it was taking the
 place of course, was completed by a degeneracy of scalar
 fields after all, what led to their mixing and breaking down
 on classes of an equivalence. It is the degeneracy of the
 primary scalar fields and hence their ability to interfere
 that allows to use the different vector fiber bundles, with
 the base being the space-time manifold. As a result the
 fundamental (uncompound) scalar particles are absent at the
 modern stage of the Universe development, of course, if we
 shall not take into account the hypothetical particles
 including Higgs' bosons.

 So the rank of the density matrix~$\rho (x)$ will be equal to 4
 and the spinors fields are becoming fundamental ones, being
 differentiated by a polarization, which is connected with
 CPT-parity. We shall use both of these varieties of the
 fields~$\Psi (x)$, noting them as
\begin{equation}
\label{1}
 \Psi = \left(\begin{array}{c}
 \psi_L\\
 \psi_R
 \end{array}  \right) .
\end{equation}
 Besides we shall double a number of components,
 using the operator $r$ of the CPT-parity in the form
\begin{equation}
\label{2}
 \psi_L = \frac12 (I - r) \psi, \quad
 \psi_R = \frac12 (I + r) \psi, \quad
 r^+ = r, \quad  r r = I
\end{equation}
 taking into account of course, that~$r$ is equal
 to~$\gamma_5 = - i \gamma_1 \gamma_2 \gamma_3 \gamma_4$
 ($i^2 = - 1$, $I$~is the unit matrix,
 and~$\gamma_1, \gamma_2, \gamma_3, \gamma_4$ are the Dirac's
 matrices) in the standard theory of quantum
 electrodynamics~(QED). But we want to obtain (but do not
 postulate) Dirac's matrices in the framework of the gauge
 theory,  therefore the operator~$r$ will not given the
 concrete expression for the present. By this we shall cosider
 that the operator of the space inversion is acting
 as~$P(\Psi ) = K\Psi$.
 Specifically the~$8\times 8$-matrix~$K$ may have the form
\begin{equation}
\label{3}
 K = \left(\begin{array}{cc}
 k&0\\
 0&k
 \end{array}\right), \quad
 k^+ = k,\quad k k = I,\quad k r + r k = 0,
\end{equation}
 so that the condition~$\overline{\Psi} \Psi = 0$ (the
 unobservability condition of the noninteraction fields~$\Psi (x)$)
 fulfiled for the noninteracton ($T_a = 0$) fields.

 Note the Lagrangian~${\cal L}$ in the form
\begin{equation}
\label{4}
 {\cal L} = \overline{D\Psi} D\Psi
\end{equation}
where
\begin{equation}
\label{5}
 D\Psi = \Phi^i (\partial_i\Psi - T_i \Psi )
\end{equation}
 and spread the gauge fields~$\Phi^i (x)$.
 Taking into account the mixing of two primary
 scalar fields, it can say that the Lee local loop~$G_r$
 contains the group SU(2)  as subloop, which has to be displaied
 by a choice of the matrices~$T_i$, if only as follows~\cite{K4}
\begin{equation}
\label{6}
 T_A = - T^+_A = \frac{i}{2}
 \Sigma_A \otimes \left(\begin{array}{cc}
                        I - k & 0 \\
                        0 & I + k
                        \end{array}\right),
\end{equation}
\begin{equation}
\label{7}
 T_4 = - T^+_4 = \frac{i}{2}
 k \otimes \left(\begin{array}{cc}
                 I - 3k & 0 \\
                 0 & I + 3k
                 \end{array}\right),
\end{equation}
 ($A,B = 1,2,3$). Defining the Dirac's matrices~$\gamma_i$ as
\begin{equation}
\label{8}
 \gamma_4 = k, \quad \gamma_A k r =
                       \Sigma_A = \left(\begin{array}{cc}
                                          \sigma_A & 0 \\
                                            0 & \sigma_A
                      \end{array}\right),
\end{equation}
 ($\sigma_A$ are the Pauli's matrices) it can obtain the
 Lagrangian~${\cal L}$ of the neutrinos fields in the standard form
\begin{equation}
\label{9}
 {\cal L} = -
 \frac i 2 [\eta^{AB} (\partial_A\overline{\psi}\gamma_B\psi
 - \overline{\psi} \gamma_A \partial_B\psi ) -
 \partial_4\overline{\psi} \gamma_4 \psi +
 \overline{\psi} \gamma_4 \partial_4\psi )]
\end{equation}

 So, having the neutrinos Universe and taking account of the
 Fermi-Dirac statistics we can recollect about the Saharov's
 hipothesis~\cite{sah}  using the idea of the metrical vacuum
 elasticity for the explanation of the gravitational interactions.
 But the main idea is it now for us what the normal matter (not
 neutrinos) acts as the Brownians by the help of which it can
 make the attempt to estimate the statistics characterization of
 the Universe neutrinos background. In the capacity of one from
 such indicator we offer to use the particles masses whiches
 connect with the scattering cross-section of the neutrinos.
 Note in tie with it, what we can ignore the photon collision
 radiation by the neutrinos scattering on the hadrons whiches
 are the quark resonator because of the existence of the additional
 degree of freedom in comparison with the electron. Exactly the
 resonance scattering causes to a gain in the hadrons masses by a
 factor of~$10^3$ in comparison with the electron mass. (The great
 spread of the hadrons masses depend on the form of the collective
 quark oscilations in the hadrons resonator.)

\end{document}